\documentclass[11pt,twoside,letterpaper]{article} 
\usepackage{times,fancyhdr}
\usepackage{graphicx}
\usepackage{epsfig}
\usepackage{epstopdf}
\usepackage{psfrag}
\usepackage{amssymb}

\makeatletter 
\def\cleardoublepage{\clearpage\if@twoside \ifodd\c@page\else%
    \hbox{}%
    \thispagestyle{empty}%
    \newpage%
    \if@twocolumn\hbox{}\newpage\fi\fi\fi} 
\makeatother

\def\figurename{Figure}
\makeatletter
\renewcommand{\fnum@figure}[1]{\figurename~\thefigure.}
\makeatother

\def\tablename{Table}
\makeatletter
\renewcommand{\fnum@table}[1]{\tablename~\thetable.}
\makeatother

\begin{document}
\title{
{\begin{flushleft}
\vskip 0.45in
{\normalsize\bfseries\textit{Chapter~1}}
\end{flushleft}
\vskip 0.45in
\bfseries\scshape Philosophical Issues of Black Holes}}
\author{\bfseries\itshape Gustavo E. Romero\thanks{E-mail address: romero@iar-conicet.gov.ar}\\
Instituto Argentino de Radioastronomia (IAR)\\
Casilla de Correos No. 5\\
Villa Elisa 1894\\
Provincia de Buenos Aires\\
Argentina}

\date{}
\maketitle
\thispagestyle{empty}
\setcounter{page}{1}
\thispagestyle{fancy}
\fancyhead{}
\fancyhead[L]{In:  \\ 
Editor:  } 
\fancyhead[R]{ISBN: \\
\copyright~2014 Nova Science Publishers, Inc.}
\fancyfoot{}
\renewcommand{\headrulewidth}{0pt}

\begin{abstract} Black holes are extremely relativistic objects. Physical processes around them occur in a regime where the gravitational field is extremely intense. Under such conditions, our representations of space, time, gravity, and thermodynamics are pushed to their limits. In such a situation philosophical issues naturally arise. In this chapter I review some philosophical questions related to black holes. In particular, the relevance of black holes for the metaphysical dispute between presentists and eternalists, the origin of the second law of thermodynamics and its relation to black holes, the problem of information, black holes and hypercomputing, the nature of determinisim, and the breakdown of predictability in black hole space-times. I maintain that black hole physics can be used to illuminate some important problems in the border between science and philosophy, either epistemology and ontology.  
\end{abstract}
  
\noindent \textbf{PACS} 04.70.Bw, 97.60.Lf, 98.80.-k, 01.70.+w.\\
\noindent \textbf{Keywords:} Black holes, cosmology, philosophy of science.

\label{lastpage-01}


\section{The Philosophical Importance of Black Holes}

Black holes are the most extreme objects known in the universe. Our representations of physical laws reach their limits in them. The strange phenomena that occur around black holes put to the test our basic conceptions of space, time, determinism, irreversibility, information, and causality. It is then not surprising that the investigation of black holes has philosophical impact in areas as diverse as ontology, epistemology, and theory construction. In black holes, in a very definite sense, we can say that philosophy meets experiment. But, alas, philosophers have almost paid no attention to the problems raised by the existence of black holes in the real world (for a notable and solitary exception see Weingard 1979; a recent discussion of some ontological implications of black holes can be found in Romero \& P\'erez 2014).

\pagestyle{fancy}
\fancyhead{}
\fancyhead[EC]{Gustavo E. Romero}
\fancyhead[EL,OR]{\thepage}
\fancyhead[OC]{Philosophical Issues of Black Holes }
\fancyfoot{}
\renewcommand\headrulewidth{0.5pt} 

The purpose of this chapter is to palliate this omission and to provide a survey of some important philosophical issues related to black holes. I do not purport to deliver an exhaustive study; such a task would demand a whole book devoted to the topic. Rather, I would like to set path for future research, calling the attention to some specific problems. 

In the next section I introduce the concept of a black hole. I do this from a space-time point of view, without connection to Newtonian analogies. Black holes are not black stars; they are fully relativistic objects and can be understood only from a relativistic perspective. Hence, I start saying a few things about space-time and relativity. 

In the remaining sections of the chapter I present and discuss several philosophical issues raised by the existence and properties of black holes. In particular, I discuss what happens with determinism and predictability in black holes space-times, the implications of the existence of black holes for ontological views of time and the nature of reality, the role of black holes in the irreversibility we observe in the universe, issues related to information and whether it can be destroyed in black holes, the apparent breakdown of causality inside black holes, and, finally, the role played, if any, by black holes in the future of the universe.

\section{What is a Black Hole?}\label{Sect2}

A black hole is a region of space-time, so I start introducing the concept of space-time (Minkowski 1908). \\

{\bf Definition.}  {\em Space-time is the emergent of the ontological composition of all events}. \\

Events can be considered as primitives or can be derived from things as changes in their properties if things are taken as ontologically prior. Both representations are equivalent since things can be construed as bundles of events (Romero 2013b). Since composition is not a formal operation but an ontological one\footnote{For instance, a human body is composed of cells, but is not just a mere collection of cells since it has emergent properties and specific functions far more complex than those of the individual components.}, space-time is neither a concept nor an abstraction, but an emergent entity. As any entity, space-time can be represented by a concept. The usual representation of space-time is given by a 4-dimensional real manifold $E$ equipped with a metric field $g_{ab}$:

$$  
{\rm ST}\hat{=}\left\langle E, g_{ab}\right\rangle.
$$

It is important to stress that space-time {\sl is not} a manifold (i.e. a mathematical construct) but the ``totality'' of events. A specific model of space-time requires the specification of the source of the metric field. This is done through another field, called the ``energy-momentum'' tensor field $T_{ab}$. Hence, a model of space-time is:

$$  
M_{\rm ST}=\left\langle E, g_{ab}, T_{ab}\right\rangle.
$$

The relation between these two tensor fields is given by field equations, which represent a basic physical law. The metric field specifies the geometry of space-time. The energy-momentum field represents the potential of change (i.e. of event generation) in space-time. 

All this can be cast into in the following axioms (Romero 2014b)\footnote{I distinguish purely syntactic from semantic axioms. The former establish relations between symbols and formal concepts. The latter, relations between concepts and elements of the reality.}.\\

${\rm P1 - Syntactic}.$ The set $E$ is a $C^{\infty}$ differentiable, 4-dimensional, real pseudo-Riemannian manifold.\\

${\rm P2 - Syntactic}. $ The metric structure of $E$ is given by a tensor field of rank 2, $g_{ab}$, in such a way that the differential distance $ds$ between two events is: $$ds^{2}=g_{ab} dx^{a} dx^{b}.$$

${\rm P3 - Syntactic}.$ The tangent space of $E$ at any point is Minkowskian, i.e. its metric is given by a symmetric tensor $\eta_{ab}$ of rank 2 and trace $-2$.\\

${\rm P4 - Syntactic}.$ The metric of $E$ is determined by a rank 2 tensor field $T_{ab}$ through the following field equations:

\begin{equation}
G_{ab}-g_{ab}\Lambda=\kappa T_{ab}, \label{Eq-Einstein} 
\end{equation}
where $G_{ab}$ is a second rank tensor whose components are functions of the second derivatives of the metric. Both $\Lambda$ and $\kappa$ are constants.\\

${\rm P5 - Semantic}.$ The elements of $E$ represent physical events.\\

${\rm P6 - Semantic}.$ Space-time is represented by an ordered pair $\left\langle E, \; g_{ab}\right\rangle$: $${\rm ST}\hat{=}\left\langle E, g_{ab}\right\rangle.$$

${\rm P7 - Semantic}.$ There is a non-geometrical field represented by a 2-rank tensor field $T_{ab}$ on the manifold E.\\ 

${\rm P8 - Semantic}.$ A specific model of space-time is given by: $$M_{{\rm ST}}=\left\langle E, g_{ab}, T_{ab}\right\rangle.$$

So far no mention has been made of the gravitational field. The sketched theory is purely ontological, and hence, cannot be yet identified with General Relativity.  To formulate the field equations we introduce the Einstein tensor:
\begin{equation}
	G_{ab}\equiv R_{ab}-\frac{1}{2}R g_{ab},
\end{equation}
where $R_{ab}$ is the Ricci tensor formed from second derivatives of the metric and $R\equiv g^{ab}R_{ab}$ is the Ricci scalar. The geodesic equations for a test particle free in the gravitational field are:
\begin{equation}
	\frac{d^{2}x^{a}}{d\lambda^{2}}+ \Gamma^{a}_{bc}\frac{dx^{b}}{d\lambda}\frac{dx^{c}}{d\lambda},
\end{equation}
with $\lambda$ an affine parameter and $\Gamma^{a}_{bc}$ the affine connection, given by:
\begin{equation}
\Gamma^{a}_{bc}=\frac{1}{2}g^{ad}(\partial_{b}g_{cd}+\partial_{c}g_{bd}-\partial_{d}g_{bc}).	
\end{equation}
  
The affine connection is not a tensor, but can be used to build a tensor that is directly associated with the curvature of space-time: the Riemann tensor. The form of the Riemann tensor for an affine-connected manifold can be obtained through a coordinate transformation 
${x^{a}\rightarrow {\bar{x}^{a}}}$ that makes the affine connection to vanish everywhere, i.e.

\begin{equation}
	\bar{\Gamma}^{a}_{bc}(\bar{x})=0, \;\;\; \forall\; \bar{x},\; a,\;b,\; c.
\end{equation}

\vspace{0.2cm}

\noindent The coordinate system ${\bar{x}^{a}}$ exists if

\begin{equation}
\Gamma^{a}_{bd, c}-\Gamma^{a}_{bc, d} + \Gamma^{a}_{ec}\,\Gamma^{e}_{bd} - \Gamma^{a}_{de}\,\Gamma^{e}_{bc}=0
\label{R}
\end{equation}

\noindent for the affine connection $\Gamma^{a}_{bc}({x})$. The left hand side of Eq. (\ref{R}) is the Riemann tensor:
\begin{equation}
	R^{a}_{bcd}=\Gamma^{a}_{bd, c}-\Gamma^{a}_{bc, d} + \Gamma^{a}_{ec}\,\Gamma^{e}_{bd} - \Gamma^{a}_{de}\,\Gamma^{e}_{bc}.
\end{equation}
 
When $R^{a}_{bcd}=0$ the metric is flat, since its derivatives are zero. If \mbox{$K=R^{a}_{bcd}R^{bcd}_{a}>0$} the metric has positive curvature. Sometimes it is said that the Riemann tensor represents the gravitational field, since it only vanishes in the absence of fields. On the contrary, the affine connection can be set locally to zero by a transformation of coordinates. This fact, however, only reflects the equivalence principle: the gravitational field can be suppressed in any locally free falling system. In other words, the tangent space to the manifold that represents space-time is always Minkowskian. To determine the mathematical object of the theory that represents the gravitational field we have to consider the weak field limit of Eqs. (\ref{Eq-Einstein}). When this is done we find that the gravitational potential is identified with the metric coefficient $g_{00}\approx \eta_{00} + h_{00}$ and the coupling constant $\kappa$ is $-8\pi G/c^{4}$. If {\em the metric represents the gravitational potential}, then {\em the affine connection represents the strength of the field itself}. This is similar to what happens in electrodynamics, where the 4-vector $A^{a}$ represents the electromagnetic potential and the tensor field $F^{ab}=\partial_{a}A_{b}-\partial_{b}A_{a}$ represents the strength of the electromagnetic field. {\em The Riemann tensor, on the other hand, being formed by derivatives of the affine connection, represents the rate of change, both in space and time, of the strength of the gravitational field}.

The source of the gravitational field in Eqs. (\ref{Eq-Einstein}), the tensor field $T_{ab}$,  stands for the physical properties of material things. It represents the energy and momentum of all non-gravitational systems.  In the case of a point mass $M$ and assuming spherical symmetry, the solution of  Eqs. (\ref{Eq-Einstein}) represents a Schwarzschild black hole. 

The Schwarzschild solution for a static mass $M$ can be written in spherical coordinates $(t,\;r,\;\theta,\;\phi)$ as:
\begin{equation}
	ds^{2}= \left(1-\frac{2GM}{rc^{2}}\right) c^{2}dt^2- \left(1-\frac{2GM}{rc^{2}}\right)^{-1} dr^{2} -r^{2} (d\theta^{2}+\sin^{2}\theta d\phi^{2}).\label{Schw}
\end{equation} 
 
 The metric given by Eq. (\ref{Schw}) has some interesting properties. Let's assume that the mass $M$ is concentrated at $r=0$. There seems to be two singularities at which the metric diverges: one at $r=0$ and the other at 
\begin{equation}
r_{\rm S}=\frac{2GM}{c^{2}}.	
\end{equation}
The length $r_{\rm S}$ is known as the {\sl Schwarzschild radius} of the object of mass $M$. Usually, at normal densities, $r_{\rm S}$ is well inside the outer radius of the physical system, and the solution does not apply to the interior but only to the exterior of the object . For a point mass, the Schwarzschild radius is in the vacuum region and the entire space-time has the structure given by (\ref{Schw}). 

It is easy to see that strange things occur close to $ r_{\rm S}$. For instance, for the proper time we get:
\begin{equation}
	d\tau=\left(1-\frac{2GM}{rc^{2}}\right)^{1/2}\;dt, \label{time1}
\end{equation}
or
\begin{equation}
	dt=\left(1-\frac{2GM}{rc^{2}}\right)^{-1/2}\;d\tau, \label{time2}
\end{equation}

When $r\longrightarrow \infty$ both times agree, so $t$ is interpreted as the proper time measure from an infinite distance. As the system with proper time $\tau$ approaches to $ r_{\rm S}$, $dt$ tends to infinity according to Eq. (\ref{time2}). The object never reaches the Schwarzschild surface when seen by an infinitely distant observer. The closer the object is to the Schwarzschild radius, the slower it moves for the external observer. 

A direct consequence of the difference introduced by gravity in the local time with respect to the time at infinity is that the radiation that escapes from a given $r>r_{\rm S}$ will be redshifted when received by a distant and static observer. Since the frequency (and hence the energy) of the photon depend on the time interval, we can write, from Eq. (\ref{time2}):
\begin{equation}
	\lambda_{\infty}=\left(1-\frac{2GM}{rc^{2}}\right)^{-1/2} \lambda.
\end{equation}
Since the redshift is:
\begin{equation}
	z=\frac{\lambda_{\infty}-\lambda}{\lambda},
\end{equation}
then
\begin{equation}
	1+z=\left(1-\frac{2GM}{rc^{2}}\right)^{-1/2},
\end{equation}
and we see that when $r\longrightarrow r_{\rm S}$ the redshift becomes infinite. This means that a photon needs infinite energy to escape from inside the region determined by $r_{\rm S}$. Events that occur at $r<r_{\rm S}$ are disconnected from the rest of the universe. The surface determined by $r=r_{\rm S}$ is an {\sl event horizon}. Whatever crosses the event horizon will never return. This is the origin of the expression ``black hole'', introduced by John A. Wheeler in the mid 1960s. The black hole is the region of space-time inside the event horizon.

 According to Eq. (\ref{Schw}), there is a divergence at $r=r_{\rm S}$. The metric coefficients, however, can be made regular by a change of coordinates. For instance we can consider Eddington-Finkelstein coordinates. Let us define a new radial coordinate $r_{*}$ such that radial null rays satisfy $d(c t\pm r_{*})=0$. Using Eq. (\ref{Schw}) we can show that:
$$r_{*}=r +  \frac{2GM}{c^{2}} \log \left|\frac{r-2GM/c^{2}}{2GM/c^{2}}\right|.$$
Then, we introduce: 
$$v=ct+r_{*}.$$ The new coordinate $v$ can be used as a time coordinate replacing $t$ in Eq. (\ref{Schw}). This yields: $$ ds^{2}=\left(1-\frac{2GM}{rc^{2}}\right) (c^2dt^{2}-dr^{2}_{*})-r^{2} d\Omega^{2}$$ or
\begin{equation}
	ds^{2}=\left(1-\frac{2GM}{rc^{2}}\right) dv^{2}-2drdv -r^{2} d\Omega^{2}, \label{EF}
\end{equation}
where $$d\Omega^{2}=d\theta^{2}+\sin^{2} \theta d\phi^{2}.$$


\begin{figure}
\centering
 \includegraphics[scale=0.9, width=7cm ]{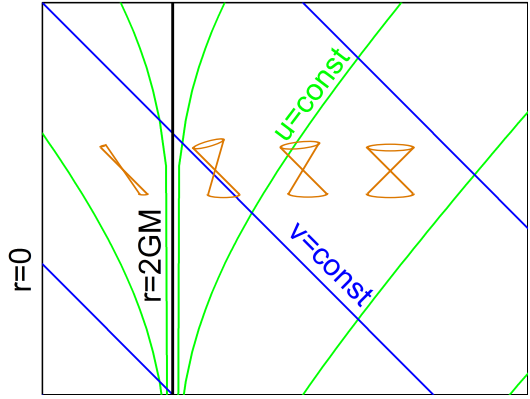}
  \caption{Space-time diagram in Eddington-Finkelstein coordinates showing the light cones close to and inside a black hole. Here, $r=2GM=r_{\rm S}$ is the Schwarzschild radius where the event horizon is located (units $c=1$).} \label{falling}
\end{figure}

Notice that in Eq. (\ref{EF}) the metric is non-singular at $r=2GM/c^{2}$. The only real singularity is at $r=0$, since there the Riemann tensor diverges. In order to plot the space-time in a $(t,\; r)$-plane, we can introduce a new time coordinate $t_{*}=v-r$. From the metric (\ref{EF}) or from Fig. \ref{falling} we see that the line $r=r_{\rm S}$, $\theta=$constant, and $\phi=$ constant is a null ray, and hence, the surface at $r=r_{\rm S}$ is a null surface. This null surface is an event horizon because inside $r=r_{\rm S}$ all cones have $r=0$ in their future (see Figure \ref{falling}). Everything that crosses the event horizon will end at the singularity. This is the inescapable fate for everything inside a Schwarzschild black hole. There is no way to avoid it: in the future of every event inside the event horizon is the singularity. However, that no signal coming from the center of the black hole can reach a falling observer, since the singularity is always in the future, and a signal can arrive only from the past.  A falling observer will never see the singularity.

Many coordinates systems can be used to describe black holes. For this reason, it is convenient to provide a definition of a black hole that is independent of the choice of coordinates. First, I will introduce some preliminary useful definitions (e.g. Hawking \& Ellis 1973, Wald 1984).\\

{\bf Definition.} {\sl A causal curve \index{causal curve} in a space-time $(M,\; g_{ab})$ is a curve that is non space-like, that is, piecewise either time-like or null (light-like).} \\

We say that a given space-time $(M, \; g_{ab})$ is {\em time-orientable} \index{space-time, time-orientable}if we can define over $M$ a smooth non-vanishing time-like vector field.\\

{\bf Definition.} {\sl If $(M,\; g_{ab})$ is a time-orientable space-time, then $\forall p\in M$, the causal future of $p$, denoted $J^{+}(p)$, is defined by:
\begin{equation}
	J^{+}(p)\equiv \left\{ q \in M | \exists \;a\; future-directed \; causal \; curve \; from  \; p \; to \; q \right\}.
\end{equation}
} 

Similarly,\\

{\bf Definition.}  {\sl If $(M,\; g_{ab})$ is a time-orientable space-time, then $\forall p\in M$, the causal past of $p$, denoted $J^{-}(p)$, is defined by}:
\begin{equation}
	J^{-}(p)\equiv \left\{ q \in M | \exists \;a\; past-directed \; causal \; curve \; from  \; p \; to \; q \right\}.
\end{equation}

The causal future \index{causal future} and past \index{causal past} of any set $S\subset M$ are given by:
\begin{equation}
	J^{+}(S) = \bigcup_{p\in S} J^{+}(p)
\end{equation}

and,
\begin{equation}
	J^{-}(S) = \bigcup_{p\in S} J^{-}(p).
\end{equation}

A set $S$ is said {\it achronal} if no two points of $S$ are time-like related. A Cauchy surface is an achronal surface such that every non space-like curve in $M$ crosses it once, and only once, $S$. A space-time $(M, g_{ab})$ is {\it globally hyperbolic} iff it admits a space-like hypersurface $S\subset M$ which is a Cauchy surface for $M$. 

Causal relations are invariant under conformal transformations of the metric. In this way, the space-times $(M, g_{ab})$ and  $(M, \widetilde{g}_{ab})$, where $\widetilde{g}_{ab}=\Omega^{2}g_{ab}$, with $\Omega$ a non-zero $C^{r}$ function, have the same causal structure. 

Let us now consider a space-time where all null geodesics that start in a region $\cal{J}^{-}$ end at $\cal{J}^{+}$. Then, such a space-time, $(M,\; g_{ab})$, is said to contain a {\it black hole} if $M$ {\it is not} contained in $J^{-}({\cal{J^+}})$. In other words, there is a region from where no null geodesic can reach the {\it asymptotic flat}\footnote{Asymptotic flatness is a property of the geometry of space-time which means that in appropriate coordinates, the limit of the metric at infinity approaches the metric of the flat (Minkowskian) space-time.} future space-time, or, equivalently, there is a region of $M$ that is causally disconnected from the global future.  The {\it black hole region}, $BH$, of such space-time is $BH=[M-J^{-}({\cal{J^+}})]$, and the boundary of $BH$ in $M$, $H=J^{-}({\cal{J^+}}) \bigcap M$, is the {\it event horizon} \index{event horizon}. 

Notice that a black hole is conceived as a space-time {\it region}, i.e. what characterises the black hole is its metric and, consequently, its curvature. What is peculiar of this space-time region is that it is causally disconnected from the rest of the space-time: no events in this region can make any influence on events outside the region. Hence the name of the boundary, event horizon: events inside the black hole are separated from events in the global external future of space-time. The events in the black hole, nonetheless, as all events, are causally determined by past events. A black hole does not represent a breakdown of classical causality. 

\begin{figure}
\centering
 \includegraphics[scale=1.2, width=10cm ]{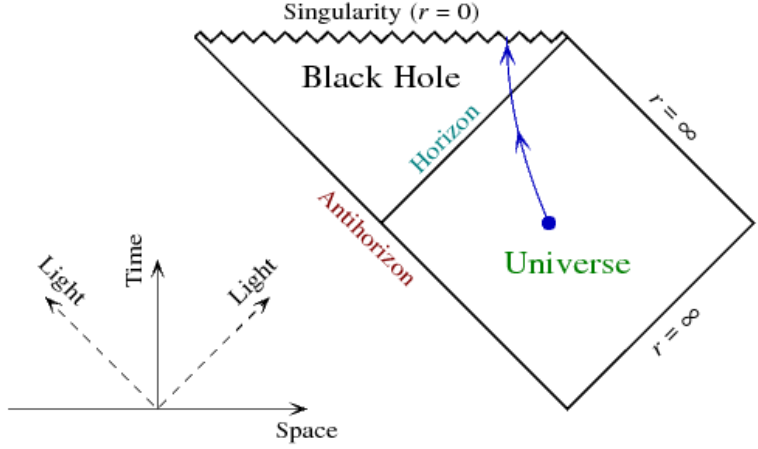}
  \caption{Carter-Penrose diagram of a Schwarzschild black hole.} \label{C-P}
\end{figure}

A useful representation of a black hole is given by a Carter-Penrose diagram. This is a two-dimensional diagram that captures the causal relations between different points in space-time. It is an extension of a Minkowski diagram where the vertical dimension represents time, and the horizontal dimension represents space, and slanted lines at an angle of $45^{\circ}$ correspond to light rays. The main difference with a Minkowski diagram (light cone) is that, locally, the metric on a Carter-Penrose diagram is conformally equivalent\footnote{I remind that two geometries are conformally equivalent if there exists a conformal transformation (an angle-preserving transformation) that maps one geometry to the other. More generally, two Riemannian metrics on a manifold $M$ are conformally equivalent if one is obtained from the other through multiplication by a function on $M$.} to the actual metric in space-time. The conformal factor is chosen such that the entire infinite space-time is transformed into a Carter-Penrose diagram of finite size. For spherically symmetric space-times, every point in the diagram corresponds to a 2-sphere. In Figure \ref{C-P}, I show a Carter-Penrose diagram of a Schwarzschild space-time.


From the Carter-Penrose diagram, it is clear that there is no time-like curve that starting from the interior region of the black hole can reach the conformally flat future infinity. All curves in this region can only end in the singularity.

Schwarzschild black holes are spherically symmetric, non-rotating objects. All known astrophysical systems have some angular momentum. In particular, since black holes of stellar mass are expected to result from the collapse of massive stars, they should be rapidly rotating objects due to the momentum conservation. The metric of a rotating mass in vacuum is the Kerr metric. For a rotating body of mass $M$ and angular momentum per unit mass $a$, this metric can be written as: 
\begin{eqnarray} ds^2&=&g_{tt}dt^2+2g_{t\phi}dtd\phi-g_{\phi\phi}d\phi^2-\Sigma\Delta^{-1}dr^2-\Sigma d\theta^2\label{Kerr}\\ g_{tt}&=& (c^2-2GMr\Sigma^{-1})
\label{Kerr1}\\ g_{t\phi}&=&2GMac^{-2}\Sigma^{-1}r\sin^2\theta\\ g_{\phi\phi}&=&[(r^2+a^2c^{-2})^2 -a^2c^{-2}\Delta \sin^2\theta ]\Sigma^{-1}\sin^2\theta\\\Sigma&\equiv& r^2+a^2 c^{-2}\cos^2\theta\\\Delta&\equiv& r^2-2GMc^{-2}r+a^2c^{-2}.
\label{Kerr2}\end{eqnarray} 

This is the Kerr metric in Boyer-Lindquist coordinates $(t,\;r,\;\theta,\;\phi)$. The metric reduces to the Schwarzschild metric for $a=0$. In Boyer-Lindquist coordinates the metric is approximately Lorentzian at infinity.

The element $g_{t\phi}$ no longer vanishes. Even at infinity this element remains (hence I wrote {\sl approximately} Lorentzian above). The Kerr parameter $a c^{-1}$ has dimensions of length. The larger the ratio of this scale to $GMc^{-2}$ (the {\it spin\ parameter\/} $a_*\equiv ac/GM$), the more aspherical the metric. Schwarzschild's black hole is the special case of Kerr's for $a=0$. Notice that, with the adopted conventions, the angular momentum $J$ is related to the parameter $a$ by:
\begin{equation}
	J=Ma.
\end{equation}

Just as the Schwarzschild solution is the unique static vacuum solution of Eqs. (\ref{einstein}) (a result called Israel's theorem), the Kerr metric is the unique stationary axisymmetric vacuum solution (Carter-Robinson theorem).  
  
The horizon, the surface which cannot be crossed outwards, is determined by the condition $g_{rr}\rightarrow\infty$ ($\Delta=0$).  It lies at $r=r^{\rm out}_{\rm h}$ where
\begin{equation}  
r^{\rm out}_{\rm h} \equiv GMc^{-2}+[(GMc^{-2})^2-a^2c^{-2}]^{1/2}.\label{rh}
\end{equation} 
Indeed, the track $r=r^{\rm out}_{\rm h}$, $\theta=$ constant with $d\phi/d\tau=a(r_{\rm h}^2+a^2)^{-1}\, dt/d\tau$ has $ds=0$ (it represents a photon circling azimuthally {\it on\/} the horizon, as opposed to hovering at it).  Hence the surface $r=r^{\rm out}_{\rm h}$ is tangent to the local light cone.  Because of the square root in Eq. (\ref{rh}), the horizon is well defined only for $a_*= ac/GM \leq 1$. An {\sl extreme} (i.e. maximally rotating) Kerr black hole has a spin parameter $a_{*}=1$. Notice that for $(GMc^{-2})^2-a^2c^{-2}>0$ we have actually two horizons. The second, the {\sl inner} horizon, is located at:
\begin{equation}  
r^{\rm inn}_{\rm h} \equiv GMc^{-2}-[(GMc^{-2})^2-a^2c^{-2}]^{1/2}.\label{rh_inn}
\end{equation}
This horizon is not seen by an external observer, but it hides the singularity to any observer that has already crossed $r_{h}$ and is separated from the rest of the universe. For $a=0$, $r^{\rm inn}_{\rm h}=0$ and $r^{\rm out}_{\rm h} =r_{\rm S}$. The case $(GMc^{-2})^2-a^2c^{-2}<0$ corresponds to no horizons and it is thought to be unphysical. 

If a particle initially falls radially with no angular momentum from infinity to the black hole, it gains angular motion during the infall. The angular velocity as seen from a distant observer is:
\begin{equation}
	\Omega (r,\; \theta)= \frac{d\phi}{dt}=\frac{(2GM/c^{2})a r}{(r^{2}+a^{2}c^{-2})^{2}-a^{2}c^{-2}\Delta\sin ^{2}\theta}.
	\label{eq:ang_velocity_kerr_BH}
\end{equation}

A particle falling into the black hole from infinite will acquire angular velocity in the direction of the spin of the black hole. As the black hole is approached, the particle will find an increasing tendency to get carried away in the same sense in which the black hole is rotating. To keep the particle stationary with respect to the distant stars, it will be necessary to apply a force against this tendency. The closer the particle will be to the black hole, the stronger the force. At a point $r_{\rm e}$ it becomes impossible to counteract the rotational sweeping force. The particle is in a kind of space-time maelstrom. The surface determined by $r_{\rm e}$ is the {\sl static limit}: from there in, you cannot avoid to rotate. Space-time is rotating here in such a way that you cannot do anything in order to not co-rotate with it. You can still escape from the black hole, since the outer event horizon has not been crossed, but rotation is inescapable. The region between the static limit and the event horizon is called the {\sl ergosphere}. The ergosphere is not spherical but its shape changes with the latitude $\theta$. It can be determined through the condition $g_{tt}=0$. If we consider a stationary particle, $r=$ constant, $\theta=$ constant, and $\phi=$ constant. Then:
\begin{equation}
c^{2}=g_{tt}\left(\frac{dt}{d\tau}\right)^{2}.	\label{sl}
\end{equation}
When $g_{tt}\leq 0$ this condition cannot be fulfilled, and hence a massive particle cannot be stationary inside the surface defined by $g_{tt}=0$. For photons, since $ds=cd\tau=0$, the condition is satisfied at the surface. Solving $g_{tt}=0$ we obtain the shape of the ergosphere: 
\begin{equation}
	r_{\rm e}=\frac{GM}{c^{2}}+\frac{1}{c^{2}}\left(G^{2}M^{2}-a^{2}c^{2}\cos^{2}\theta\right)^{1/2}.
\end{equation}

 The static limit lies outside the horizon except at the poles where both surfaces coincide. The phenomenon of ``frame dragging''' is common to all axially symmetric metrics with $d_{t\phi}\neq 0$.   

An essential singularity occurs when $g_{tt}\rightarrow\infty$. This happens if $\Sigma=0$. This condition implies:
\begin{equation}
	r^2+a^2 c^{-2}\cos^2\theta=0.
\end{equation}
Such a condition is fulfilled only by $r=0$ and $\theta=\frac{\pi}{2}$. This translates in Cartesian coordinates to\footnote{The relation with Boyer-Lindquist coordinates is $x=\sqrt{r^{2}+a^{2}c^{-2}}\sin \theta \cos\phi$, $y=\sqrt{r^{2}+a^{2}c^{-2}}\sin \theta \sin\phi$, $z=r \cos \theta.$ }:
\begin{equation}
	x^{2}+y^{2}=a^{2}c^{-2}\;\;\;\;{\rm and}\;\;\;\;z=0.
\end{equation}
 The singularity is a ring of radius $ac^{-1}$ on the equatorial plane. If $a=0$, then Schwarzschild's point-like singularity is recovered. If $a\neq0$ the singularity is not necessarily in the future of all events at $r<r^{\rm inn}_{\rm h}$: the singularity can be avoided by some geodesics. 

\begin{figure}
\centering
 \includegraphics[scale=0.7, width=7cm ]{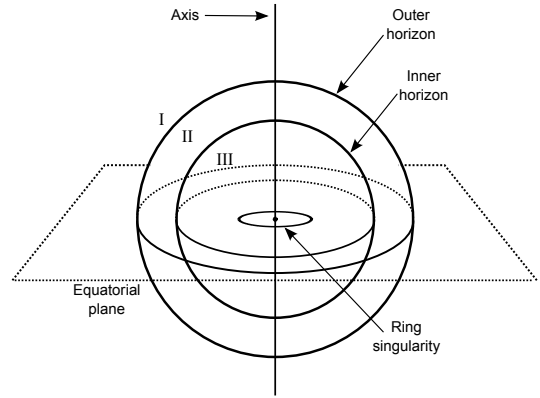}
  \caption{Sketch of a Kerr black hole, with its two horizons and the ring singulatrity. } \label{KBH}
\end{figure}

A sketch of a Kerr black hole is shown in Figure \ref{KBH}. 

Non-vacuum solutions of both spherically symmetric and rotating black holes exists, but since they are thought to be of no astrophysical importance, I do not discuss them here (the interested reader can see Romero \& Vila 2014 and Punsly 2001).

\section{Determinism and Predictability in Black Hole Spacetimes}\label{Sect3}

Determinism is a metaphysical doctrine about the nature of the world. It is an ontological assumption: the assumption that all events are given. It can be traced to Parmenides and his ``what is, is'' (Romero 2012). It is important to emphasise that determinism does not require causality and does not imply predictability. 
Predictability is a property of our theories about the world, not a property of the world itself. 

The confusion between determinism and predictability can be traced to Pierre-Simon Laplace and his {\em Philosophical Essay on Probabilities}:

\begin{quotation}
We may regard the present state of the Universe as the effect of its past and the cause of its future. An intellect which at a certain moment would know all forces that set nature in motion, and all positions of all items of which nature is composed, if this intellect were also vast enough to submit these data to analysis, it would embrace in a single formula the movements of the greatest bodies of the Universe and those of the tiniest atom; for such an intellect nothing would be uncertain and the future just like the past would be present before its eyes.
\end{quotation}

According to Laplace, every state of the Universe is determined by a set of initial conditions and the laws of physics. Since the laws are represented by differential equations and there are theorems for the existence and uniqueness of solutions, determinism implies predictability. Theorems apply, however, only to mathematical objects, not to reality. The world is not mathematical, just some of our representations of it are mathematical. The existence of solutions to some equations that represent physical laws does not imply physical existence. Physical existence is independent of our conceptions. Moreover, even in Newtonian space-times there are Cauchy horizons (Earman 1986). These are hypersurfaces from where, even the in case of a complete specification of initial data, the solutions of dynamical equations cannot predict all future events. This arises because of the absence of an upper bound on the velocities of moving objects in the Newtonian physics. For instance, consider the trajectory of an object that is accelerated in such a way that its velocity becomes in effect infinite in a finite time. This object will be disconnected from all later times. 

General Relativity assumes the existence of all events represented by a manifold (see the axiomatic system presented  in Section \ref{Sect2}). Hence, it is a deterministic theory from an ontological point of view.
The Cauchy problem, however, cannot always be solved in General Relativity. Cauchy horizons naturally appear in many solutions of Einstein field equations, and in particular, in those of rotating black holes.  The inner horizons of both Kerr and Kerr-Newman black holes are Cauchy surfaces: it is impossible to predict the evolution of any physical system in the interior region from the specification of the initial conditions over the horizon and the Einstein equations.   Although the manifold is fixed, we cannot always describe it from limited knowledge. General Relativity is an example of a physical theory that can be ontologically deterministic but nonetheless epistemologically underdetermined.  

I remark that the existence of singular space-time models $$M^{\rm sing}_{{\rm ST}}=\left\langle E, g_{ab}, T_{ab}\right\rangle$$ does not imply a breakdown of the ontological determinacy of the theory. Singularities, certainly, imply a failure in the predictability, but they are not elements of space-time itself. I will elaborate more about this in Section \ref{Sect7}

The fact that there exist irreversible processes in the universe implies that space-time is globally asymmetric. The laws that constrain the space-state of physical things, and therefore their potential to change, however, are invariant under time reversal.  Black holes might play a crucial role to link the the global structure of space-time with the local irreversibility expressed by the Second Law of Thermodynamics. I turn now to this problem. 

\section{Black Holes and the Second Law of Thermodynamics}\label{Sect4}

The Second Law of Thermodynamics states that {\em the entropy of a closed system never decreases}. If entropy is denoted by $S$, this law reads:
\begin{equation}
	\frac{dS}{dt}\geq 0.
\end{equation}

In the 1870s, Ludwig Boltzmann argued that the effect of randomly moving gas molecules is to ensure that the entropy of a gas would increase, until it reaches its maximum possible value. This is his famous {\em H-theorem}. Boltzmann was able to show that macroscopic distributions of great inhomogeneity (i.e. of high order or low entropy) are formed from relatively few microstate arrangements of molecules, and were, consequently, relatively improbable. Since physical systems do not tend to go into states that are less probable than the states they are in, it follows that any system would evolve toward the macrostate that is consistent with the larger number of microstates. The number of microstates and the entropy of the system are related by the fundamental formula:
\begin{equation}
	S= k \ln W,
\end{equation}
where $k=10^{-23}$ JK$^{-1}$ is Boltzmann's constant and $W$ is the volume of the phase-space that corresponds to the macrostate of entropy $S$. 

More than twenty years after the publication of Boltzmann's fundamental papers on kinetic theory, it was pointed out by Burbury (1894, 1895) that the source of asymmetry in the H-theorem is the implicit assumption that the motions of the gas molecules are independent before they collide and not afterwards. This essentially means that the entropy increase is a consequence of the {\em initial conditions} imposed upon the state of the system. Boltzmann's response was: 

\begin{quotation}
There must then be in the universe, 
which is in thermal equilibrium as a 
whole and therefore dead, here and 
there, relatively small regions of the 
size of our  world, which during the 
relatively short time of eons deviate 
significantly from thermal equilibrium.  
Among these worlds the state probability 
increases as often as it decreases.\\

\begin{flushright}
Boltzmann (1895).
\end{flushright}

\end{quotation} 

As noted by Price (2004): ``The low-entropy condition of our region seems to be associated entirely with a low-energy condition in our past.''

The probability of the large fluctuations required for the formation of the universe we see, on other hand, seems to be zero, as noted long ago by Eddington (1931): ``A universe containing mathematical physicists 
 at any assigned date will be in the state of 
maximum disorganisation which is not inconsistent 
with the existence of such creatures.'' Large fluctuations are rare ($P\sim \exp{-\Delta S}$); {\em extremely} large fluctuation, basically impossible. For the whole universe, $\Delta S\sim 10^{104}$ in units of $k=1$. This yields $P=0$.

In 1876, a former teacher of Boltzmann and later colleague at the University of Vienna, J. Loschmidt, noted:

\begin{quotation}
Obviously, in every arbitrary system the course of events must become retrograde when the velocities of all its elements are reversed.\\
 
\begin{flushright}
Loschmidt (1876).
\end{flushright}

\end{quotation}

In modern terminology, the laws of (Hamiltonian) mechanics are such that for every solution one can construct another solution by reversing all velocities and replacing $t$ by $-t$. Since the Boltzmann's function $H[f]$ is invariant under velocity reversal, it follows that if $H[f]$ decreases for the first solution, it will increase for the second. Accordingly, the reversibility objection is that the H-theorem cannot be a general theorem for all mechanical evolutions of the gas. More generally, the problem goes far beyond classical mechanics and encompasses our whole representation of the physical world. This is because {\em all formal representations of all fundamental laws of physics are invariant under the operation of time reversal}. Nonetheless, the evolution of all physical processes in the universe is irreversible. 

If we accept, as mentioned, that the origin of the irreversibility is not in the laws but in the initial conditions of the laws, two additional problems emerge: 1) What were exactly these initial conditions?, and 2) How the initial conditions, of global nature, can enforce, at any time and any place, the observed local irreversibility? 

The first problem is, in turn, related to the following one, once the cosmological setting is taken into account: in the past, the universe was hotter and at some point matter and radiation were in thermal equilibrium; how is this compatible with the fact that entropy has ever been increasing according to the so-called Past Hypothesis, i.e. entropy was at a minimum at some past time and has been increasing ever since?  

The standard answer to this question invokes the expansion of the universe: as the universe expanded, the maximum possible entropy increased with the size of the universe, but the actual entropy was left well behind the permitted maximum. The source of irreversibility in the Second Law of Thermodynamics is the trend of the entropy to reach the permitted maximum. According to this view, the universe actually began in a state of maximum entropy, but due to the expansion, it was still possible for the entropy to continue growing.        

The main problem with this line of thought is that is not true that the universe was in a state of maximum disorder at some early time. In fact, although locally matter and radiation might have been in thermal equilibrium, this situation occurred in a regime were the global effects of gravity cannot be ignored (Penrose 1979). Since gravity is an attractive force, and the universe was extremely smooth (i.e structureless) in early times, as indicated, for instance, by the measurements of the cosmic microwave background radiation, the gravitational field should have been quite far from equilibrium, with very low global entropy (Penrose 1979). It seems, then, that the early universe was {\em globally} out of the equilibrium, being the total entropy dominated by the entropy of the gravitational field. If we denote by $C^{2}$ a scalar formed out by contractions of the Weyl tensor, the initial condition $C^{2}\sim 0$ is required if entropy is still growing today\footnote{This is because the Weyl tensor provides a measure of the inhomogeneity of the gravitational field. See Romero, Thomas, \& P\'erez (2012) for estimates of the gravitational entropy of black holes based on the Weyl tensor.}.  

The answer to the second question posed above, namely,  `how the Second Law is locally enforced  by the initial conditions, which are of global nature?', seems to require a coupling between gravitation (of global nature) and electrodynamics (of local action). In what follows I suggest that black holes can provide the key for this coupling (for the role of cosmological horizons in this problem see Romero \& P\'erez 2011).

The electromagnetic radiation field can be described in the terms of the 4-potential $A^{\mu}$, which in the Lorentz gauge satisfies:
\begin{equation}
\partial^{b}\partial_{b}A^{a}(\vec{r},\;t)=4\pi j^{a} (\vec{r},\;t),
\end{equation}
with $c=1$ and $j^{a}$ the 4-current. The solution $A^{a}$ is a functional of the sources $j^{a}$. The retarded and advanced solutions are:

\begin{equation}
	A^{a}_{\rm ret}(\vec{r},\;t)=\int_{V_{\rm ret}}
\frac{j^{a} \left(\vec{r},\;t-\left|\vec{r}-\vec{r'}\right|\right)}{\left|\vec{r}-\vec{r'}\right|}d^{3}\vec{r'} + \int_{\partial V_{\rm ret}}
\frac{j^{a} \left(\vec{r},\;t-\left|\vec{r}-\vec{r'}\right|\right)}{\left|\vec{r}-\vec{r'}\right|}d^{3}\vec{r'}, \label{ret}
\end{equation}

\begin{equation}
	A^{a}_{\rm adv}(\vec{r},\;t)=\int_{V_{\rm adv}}
\frac{j^{a} \left(\vec{r},\;t+\left|\vec{r}-\vec{r'}\right|\right)}{\left|\vec{r}-\vec{r'}\right|}d^{3}\vec{r'} + \int_{\partial V_{\rm adv}}
\frac{j^{a} \left(\vec{r},\;t+\left|\vec{r}-\vec{r'}\right|\right)}{\left|\vec{r}-\vec{r'}\right|}d^{3}\vec{r'}. \label{adv}
\end{equation}

The two functionals of $j^{a}(\vec{r},\;t)$ are related to one another by a time reversal transformation. The solution (\ref{ret}) is contributed by sources in the past of the space-time point $p(\vec{r},\;t)$ and the solution (\ref{adv}) by  sources in the future of that point. The integrals in the second term on the right side are the surface integrals that give the contributions from i) sources outside of $V$ and ii) source-free radiation. If $V$ is the causal past and future, the surface integrals do not contribute. 

  
The linear combinations of electromagnetic solutions are also solutions, since the equations are linear and the Principle of Superposition holds. It is usual to consider only the retarded potential as physical meaningful in order to estimate the electromagnetic field at $p(\vec{r},\;t)$: $F^{ab}_{\rm ret}=\partial^{a}A^{b}_{\rm ret}-\partial^{b}A^{a}_{\rm ret}$. However, there seems to be no compelling reason for such a choice. We can adopt, for instance (in what follows I use a simplified notation),
\begin{equation}
	A^{a}(\vec{r},\;t)=\frac{1}{2}\left(\int_{J^{+}} {\rm adv\;} + \;\int_{J^{-}} {\rm ret}\right)\; dV.
\end{equation}
 
If the space-time is curved ($R^{abcd}R_{abcd}\neq 0$), the null cones that determine the causal structure will not be symmetric around the point $p$ $(\vec{r},\;t)$. In particular, the presence of event horizons can make very different the contributions from both integrals. 

Hawking's black hole area theorem (Hawking 1971) ensures that in a time-orientable space-time such that for all null vectors $k^{a}$ holds $R_{ab}k^ak^b \geq 0$, the area of the event horizons of black holes either remains the same or increases with cosmic time. More precisely:\\

{\bf Theorem.} {\sl Let $(M,\; g_{ab})$ be a time-orientable space-time such that   $R_{ab}k^ak^b\geq 0$ for all null $k^a$. Let $\Sigma_1$ and   $\Sigma_2$ be space-like Cauchy surfaces for the globally hyperbolic region of the space-time with $\Sigma_2 \subset J^+(\Sigma_1)$, and be ${\cal H}_1= H \bigcap \Sigma_1$,  ${\cal H}_2= H \bigcap \Sigma_2$, where $H$ denotes an event horizon. Then ${\cal H}_2\geq {\cal H}_1$.} \\

The fact that astrophysical black holes are always immersed in the cosmic background radiation, whose temperature is much higher than the horizon temperature,
implies that they always accrete and then, by the first law of black holes (Bardeen et al. 1973), ${\cal H}_2 > {\cal H}_1$. The total area of black holes increases with cosmic time.  The accretion should include not only photons but also charged particles. This means that the total number of charges in the past of any point $p(\vec{r},\;t)$ will be different from their number in the corresponding future. This creates a local asymmetry that can be related to the Second Law. 
    
We can introduce a vector field $L^{a}$ given by:  
   
\begin{equation}
L^{a}= \left[\int_{J^{-}} {\rm ret} - \int_{J^{+}} {\rm adv} \right]\; dV \neq 0.
\end{equation}

If $g_{ab}L^{a}T^{b}\neq0$, with $T^{b}=(1,0,0,0)$ there is a preferred direction for the Poynting flux in space-time. The Poynting flux is given by:

\begin{equation}
	\vec{S}=4\pi \vec{E} \times \vec{B}= (T^{01}_{\rm EM}, T^{02}_{\rm EM}, T^{03}_{\rm EM}),
\end{equation}
where $\vec{E}$ and $\vec{B}$ are the electric and magnetic fields and $T^{ab}_{\rm EM}$ is the electromagnetic energy-momentum tensor.

In a black hole interior the direction of the Poynting flux is toward the singularity. In an expanding, accelerating universe, it is in the global future direction. We see, then, that a time-like vector field, in a general space-time $(M, g_{ab})$, can be {\sl anisotropic}. There is a global to local relation given by the Poynting flux as determined by the curvature of space-time that indicates the direction along which events occur. Physical processes, inside a black hole, have a different orientation from outside, and the causal structure of the world is determined by the dynamics of space-time and the initial conditions. Macroscopic irreversibility\footnote{Notice that the electromagnetic flux is related with the macroscopic concept of temperature through the Stefan-Boltzmann law: $L=A\sigma_{\rm SB}T^{4}$, where $\sigma_{\rm SB}$ is the Stefan-Boltzmann constant.} and time anisotropy emerge from fundamental reversible laws. 

There is an important corollary to these conclusions. Local observations about the direction of events can provide information about global features of space-time and the existence of horizons and singularities.

\section{Time and Black Holes}\label{Sect5}

Presentism is a metaphysical thesis about what there is. It can be expressed as (e.g. Crisp 2003):
\begin{quotation}
{\it Presentism}. It is always the case that, for every $x$, $x$ is present. 
\end{quotation}
The quantification in this scheme is unrestricted, it ranges over all existents. In order to render this definition meaningful, the presentist must provide a specification of the term `present'. Crisp, in the cited paper, offers the following definition:
\begin{quotation}
{\it Present}. The mereological sum of all objects with null temporal distance. 
\end{quotation} 
The notion of temporal distance is defined loosely, but in such a way that it accords with common sense and the physical time interval between two events. From these definitions it follows that the present is a thing, not a concept. The present is the ontological aggregation of all present things. Hence, to say that `$x$ is present', actually means ``$x$ is part of the present".   

The opposite thesis of presentism is eternalism, also called four-dimensionalism. Eternalists subscribe the existence of past and future objects. The temporal distance between these objects is non-zero. The name four-dimensionalism comes form the fact that in the eternalist view, objects are extended through time, and then they have a 4-dimensional volume, with 3 spatial dimensions and 1 time dimension. There are different versions of eternalism. The reader is referred to Rea (2003) and references therein for a discussion of eternalism.

I maintain that presentism is incompatible with the existence of black holes. Let us see briefly the argument, considering, for simplicity, Schwarzschild black holes (for details, see Romero \& P\'erez 2014).  

The light cones in Schwarzschild space-time can be calculated from the metric (\ref{Schw}) imposing the null condition $ds^{2}=0$. Then: 
\begin{equation}
	\frac{dr}{dt}=\pm\left(1-\frac{2GM}{r}\right), \label{cones-Schw}
\end{equation}
where I made $c=1$. Notice that when $r\rightarrow \infty$, $dr/dt \rightarrow \pm 1$, as in Minkowski space-time. When $r\rightarrow 2GM$, $dr/dt \rightarrow 0$, and light moves along the surface $r=2GM$. The horizon is therefore a {\it null surface}. For $r<2GM$, the sign of the derivative is inverted. The inward region of $r=2GM$ is time-like for any physical system that has crossed the boundary surface. As we approach to the horizon from the flat space-time region, the light cones become thinner and thinner indicating the restriction to the possible trajectories imposed by the increasing curvature. On the inner side of the horizon the local direction of time is `inverted' in the sense that all null or time-like trajectories have in their future the singularity at the center of the black hole.

There is a very interesting consequence of all this: an observer on the horizon will have her present {\it along} the horizon. All events occurring on the horizon are simultaneous. The temporal distance from the observer at any point on the horizon to any event occurring on the horizon is zero (the observer is on a null surface $ds=0$ so the proper time interval is necessarily zero\footnote{Notice that this can never occur in Minkowski space-time, since there only photons can exist on a null surface. The black hole horizon, a null surface, can be crossed, on the contrary, by massive particles.}). If the black hole has existed during the whole history of the universe, all events on the horizon during such history (for example the emission of photons on the horizon by infalling matter) are {\it present} to any observer crossing the horizon. These events are certainly not all present to an observer outside the black hole. If the outer observer is a presentist, she surely will think that some of these events do not exist because they occurred or will occur either in the remote past or the remote future. But if we accept that what there is cannot depend on the reference frame adopted for the description of the events, it seems we have an argument against presentism here. Before going further into the ontological implications, let me clarify a few physical points. 

I remark that the horizon 1) does not depend on the choice of the coordinate system adopted to describe the black hole, 2) the horizon is an absolute null surface, in the sense that this property is intrinsic and not frame-dependent, and 3) it is a non-singular surface (or `well-behaved', i.e. space-time is regular on the horizon).

In a world described by special relativity, the only way to cross a null surface is by moving faster than the speed of light. As we have seen, this is not the case in a universe with black holes. We can then argue against presentism along the following lines. \\ 

Argument $A1$:

\begin{itemize}
	\item{$P1$: There are black holes in the universe.}
	\item{$P2$: Black holes are correctly described by General Relativity.}
	\item{$P3$: Black holes have closed null surfaces (horizons).}
	
	\item{Therefore, there are closed null surfaces in the universe.}
\end{itemize}

Argument $A2$:

\begin{itemize}
	\item{$P4$: All events on a closed null surface are simultaneous with any event on the same surface.}
	\item{$P4i$: All events on the closed null surface are simultaneous with the birth of the black hole.}
	\item{$P5$: Some distant events are simultaneous with the birth of the black hole, but not with other events related to the black hole.}
	
	\item{Therefore, there are events that are simultaneous in one reference frame, and not in another.}
\end{itemize}

Simultaneity is frame-dependent. Since what there exist cannot depend on the reference frame we use to describe it, we conclude that there are non-simultaneous events. Therefore, presentism is false.

Let us see which assumptions are open to criticism by the presentist. 

An irreducible presentist might plainly reject $P1$. Although there is significant astronomical evidence supporting the existence of black holes (e.g. Camenzind 2007, Paredes 2009, Romero and Vila 2014), the very elusive nature of these objects still leaves room for some speculations like gravastars and other exotic compact objects. The price of rejecting $P1$, however, is very high: black holes are now a basic component of most mechanisms that explain extreme events in astrophysics, from quasars to the so-called gamma-ray bursts, from the formation of galaxies to the production of jets in binary systems. The presentist rejecting black holes should reformulate the bulk of contemporary high-energy astrophysics in terms of new mechanisms. In any case, $P1$ is susceptible of empirical validation through direct imagining of the super-massive black hole ``shadow'' in the center of our galaxy by sub-mm interferometric techniques in the next decade (e.g. Falcke et al. 2011). In the meanwhile, the cumulative case for the existence of black holes is overwhelming, and very few scientists would reject them on the basis of metaphysical considerations only.

The presentist might, instead, reject $P2$. After all, we {\it know} that General Relativity fails at the Planck scale. Why should it provide a correct description of black holes? The reason is that the horizon of a black hole is quite far from the region where the theory fails (the singularity). The distance, in the case of a Schwarzschild black hole, is $r_{\rm S}$. For a black hole of 10 solar masses, as the one suspected to form part of the binary system Cygnus X-1, this means $30$ km. And for the black hole in the center of the galaxy, about 12 million km. Any theory of gravitation must yield the same results as General Relativity at such distances. So, even if General Relativity is not the right theory for the classical gravitational field, the correct theory should predict the formation of black holes under the same conditions.

There is not much to do with $P4$, since it follows from the condition that defines the null surface: $ds=0$\footnote{$ds=cd\tau=0 \rightarrow d\tau=0$, where $d\tau$ is the proper temporal separation.}; similarly $P4i$ only specifies one of the events on the null surface. A presentist might refuse to identify `the present' with a null surface. After all, in Minkowskian space-time or even in a globally time-orientable pseudo-Riemannian space-time the present is usually taken as the hyperplane perpendicular to the local time. But in space-times with black holes, the horizon is not only a null surface; it is also a surface locally normal to the time direction.  In a Minkowskian space-time  the plane of the present is not coincident with a null surface. However, close to the event horizon of a black hole, things change, as indicated by Eq. (\ref{cones-Schw}). As we approach the horizon,  the null surface matches the plane of the present. On the horizon, both surfaces are exactly coincident. A presentist rejecting the identification of the present with a {\it closed} null surface on an event horizon should abandon what is perhaps her most cherished belief: the identification of `the present' with hypersurfaces that are normal to a local time-like direction.

The result mentioned above is not a consequence of any particular choice of coordinates but an intrinsic property of a black hole horizon. This statement can be easily proved. The symmetries of Schwarzschild space-time imply the existence of a preferred radial function, $r$, which serves as an affine parameter along both null directions. The gradient of this function, $r_{a}=\nabla_{a} r$ satisfies ($c=G=1$):

\begin{equation}
r^{a}r_{a}=\left(1-\frac{2M}{r}\right). \label{ra}
\end{equation}
Thus, $r^{a}$ is space-like for $r>2M$, null for $r=2M$, and time-like for $r<2M$. The 3-surface given by $r=2M$ is the horizon $H$ of the black hole in Schwarzschild space-time. From Eq. (\ref{ra}) it follows that $r^{a}r_{a}=0$ over $H$, and hence $H$ is a null surface\footnote{An interesting case is Schwarzschild space-time in the so-called Painlev\'e-Gullstrand coordinates. In these coordinates the interval reads: 
\begin{equation}
	ds^{2}=dT^{2}-\left(dr + \sqrt{\frac{2M}{r}} dT\right)^{2} - r^{2}d\Omega^{2},\label{PG}
\end{equation}
with
\begin{equation}
	T=t + 4M \left(\sqrt{\frac{2M}{r}} + \frac{1}{2} \ln \left| \frac{\sqrt{\frac{2M}{r}}-1}{\sqrt{\frac{2M}{r}}+1} \right|\right).
\end{equation}

If a presentist makes the choice of identifying the present with the surfaces of $T=$constant, from Eq. (\ref{PG}): $ds^{2}= - dr^{2} - r^{2}d\Omega^{2}$. Notice that for $r=2M$ this is the event horizon, which in turn, is a null surface. Hence, with such a choice, the presentist is considering that the event horizon is the hypersurface of the present, for all values of $T$. This choice of coordinates makes particularly clear that the usual presentist approach to define the present in general relativity self-defeats her position if space-time allows for black holes.}.  

Premise $P5$, perhaps, looks more promising for a last line of presentist defence. It might be argued that events on the horizon are not simultaneous with any event in the external universe. They are, in a very precise sense, cut off from the universe, and hence cannot be simultaneous with any distant event. Let us work out a counterexample. 

The so-called long gamma-ray bursts are thought to be the result of the implosion of a very massive and rapidly rotating star. The core of the star becomes a black hole, which accretes material from the remaining stellar crust. This produces a growth of the black hole mass and the ejection of matter from the magnetised central region in the form of relativistic jets (e.g. Woosely 1993). Approximately, one of these events occur in the universe per day. They are detected by satellites like {\it Swift} (e.g. Piran and Fan 2007), with durations of a few tens of seconds. This is the time that takes for the black hole to swallow the collapsing star. Let us consider a gamma-ray burst of, say, 10 seconds. Before these 10 seconds, the black hole did not exist for a distant observer $O1$. Afterwards, there is a black hole in the universe that will last more than the life span of any human observer. Let us now consider an observer $O2$ collapsing with the star. At some instant she will cross the null surface of the horizon. This will occur within the 10 seconds that the collapse lasts for $O1$. But for $O2$ all photons that cross the horizon are simultaneous, including those that left $O1$ long after the 10 seconds of the event and crossed the horizon after traveling a long way. For instance, photons  leaving the planet of $O1$ one million years after the gamma-ray burst, might cross the horizon, and then can interact with $O2$. So, the formation of the black hole is simultaneous with events in $O1$ and $O2$, but these very same events of $O2$ are simultaneous with events that are in the distant future of $O1$.

The reader used to work with Schwarzschild coordinates perhaps will object that $O2$ never reaches the horizon, since the approaching process takes an infinite time in a distant reference frame. This is, however, an effect of the choice of the coordinate system and the test-particle approximation (see, for instance, Hoyng 2006, p.116). If the process is represented in Eddington-Finkelstein coordinates, it takes a finite time for the whole star to disappear, as shown by the fact that the gamma-ray burst are quite short events. Accretion/ejection processes, well-documented in active galactic nuclei and microquasars (e.g. Mirabel et al. 1998) also show that the time taken to reach the horizon is finite in the asymptotically flat region of space-time.   

My conclusion is that black holes can be used to show that presentism provides a defective picture of the ontological substratum of the world.

\section{Black Holes and Information}\label{Sect6}

Black holes are often invoked in philosophical (and even physical) discussions about production and destruction of `information'. This mostly occurs in relation to the possibility hypercomputing and the application of quantum field theory to the near horizon region. I shall review both topics here.

The expression `hypercomputing' refers to the actual performance of an infinite number of operations in a finite time with the aim of calculating beyond the Turing barrier (Turing, 1936. For a definition of a Turing machine see  Hopcrof \& Ullman 1979). It has been suggested that such a hypercomputation can be performed in a Kerr space-time (N\'emeti \& David 2006, N\'emeti \& Handr\'eka 2006). The Kerr space-time belongs to the class of the so-called Malament-Hogarth (M-H) space-times. These are defined as follows (Hogarth 1994):\\

{\bf Definition.} {\sl  $(M,\; g_{ab})$ is an M-H space-time if there is a future-directed time-like
half-curve $\gamma \subset M$ and a point $p \in M$ such that $\int_{\gamma} d\tau=\infty$ and $\gamma \subset J^- (p)$.} \\

\begin{figure}[ht]
\centering
\includegraphics[scale=0.9, width=7cm ]{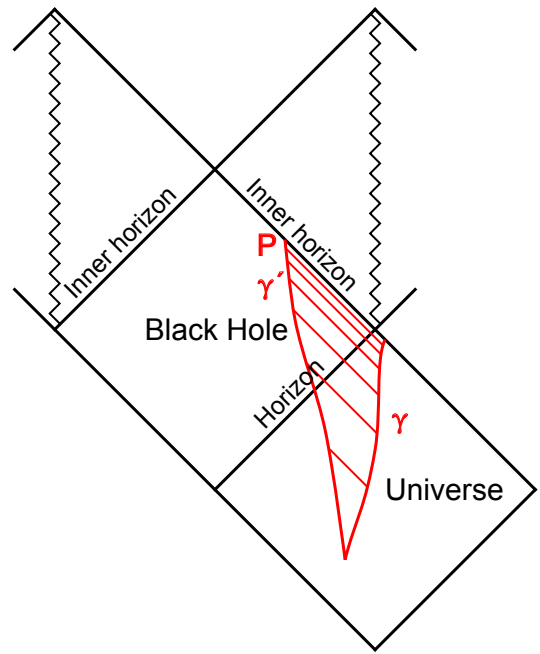}
  \caption{Carter-Penrose diagram of a Kerr black hole. The trajectories of two physical systems are indicated: $\gamma$ remains in the exterior space-time for an infinite amount of time, whereas $\gamma '$ falls into the black hole. In the time it takes the latter to reach the inner horizon, the former arrives to the conformal infinity. The lines that connect both trajectories represent signals sent from $\gamma$ to $\gamma '$.}  \label{C-P2}
\end{figure}

The curve $\gamma$ represents the world-line of some physical system. Because $\gamma$ has infinite proper time,  it may complete an infinite number of tasks. But, at every point
in $\gamma$ , it is possible to send a signal to the point $p$. This is because there always exists a curve $\gamma'$ with future endpoint $p$ which has finite proper time.  We can think of $\gamma$ as the ``sender'' and $\gamma ' $ as the ``receiver'' of a signal. In this way, the receiver may obtain knowledge of the result of an infinite number of tasks in a finite time. In a Kerr space-time this scheme can be arranged as follows. The ``sender'' is a spacecraft orbiting the Kerr black hole with a computer onboard. The ``receiver'' is a capsule ejected by the orbiter that falls into the black hole. As the capsule approaches the inner horizon it intersects more and more signals from the orbiter, which emits periodically results of the computer calculations into the black hole. By the time the capsule crosses the inner horizon it has received all signals emitted by the computer in an infinite time (assuming that both the black hole and the orbiter can exist forever). This would allow the astronauts in the capsule to get answers to questions that require beyond-Turing computation! (N\'emeti \& David 2006). The whole situation is depicted in Figure \ref{C-P2}.

There are many reasons to think that the described situation is physically impossible. I shall mention the following ones: 1) The required inner black hole structure does not correspond to an astrophysical black hole generated by gravitational collapse. In a real black hole the Cauchy horizon is expected to collapse into a (probably null) singularity due to the backscattered gravitational wave tails that enter the black hole and are blueshifted at the Cauchy horizon (see next section and Brady 1999). The instability of the Cauchy horizon seems to be a quite general feature of any realistic black hole interior model.  2) The black hole is not expected to exist during an infinite duration: it should evaporate through Hawking radiation, over very long (but always finite) time. 3) The performance of infinite operations would require an infinite amount of energy (Bunge 1977, Romero 2014). Even if the universe were infinite, a finite spacecraft cannot manipulate infinite amounts of energy. 4) If signals are periodically sent to the receiver, the blushifted electromagnetic radiation would burn the capsule by the time it crosses the Cauchy horizon. N\'emeti \& David (2006) argue that this might be circumvented by sending just one signal with the final result. This suggestion faces the problems of the actual infinite: for any moment there will always be a further moment, then, when the spaceship would send this signal? 5) The universe seems to be entering into a de Sitter phase, so particle horizons will appear and block part of the accessible space-time to the spacecraft limiting its resources.

I think that the cumulative argument is strong enough to support a {\sl hypercomputing avoidance conjeture}: the laws of physics are such that no actual hypercomputation can be performed. 

I turn now to another issue related to black holes and information: the destruction of information by black holes. This seems to be a topic of high concern for quantum field theorists, to the point that the presumed destruction of information in a black hole is called the ``black hole information paradox''. I maintain that such a paradox does not exist: black holes cannot destroy any information. The reason is that information is not a property of physical systems. It is not like the electric charge, mass, or angular momentum. Information is an attribute of {\sl languages}, and languages are constructs, i.e. elaborated fictions. To say that black holes can destroy information is like to say that they can destroy syntax. Let us review the situation in a bit more detail.    

The application of quantum field theory to the near horizon region of a black hole results in the prediction of thermal radiation (Hawking 1974). A temperature, then, can be associated with the horizon:

\begin{equation}
	T_{\rm BH}=\frac{\hbar c^{3}}{8GMk}\cong 10^{-7}{\rm K} \left(M_{\odot}\over M \right). \label{T}
\end{equation}
 We can write the entropy of the black hole as:
\begin{equation}
	S=\int \frac{dQ}{T_{\rm BH}}=\frac{kc^{3}}{4\pi \hbar G} A_{\rm BH} + \;{\rm constant}\sim 10^{77} \left(\frac{M}{M_{\odot}}\right)^{2} k \; {\rm JK}^{-1}. \label{ent}
\end{equation}

The area of a Schwarzschild black hole is:
\begin{equation}
	A_{\rm Schw}=4\pi r^{2}_{\rm Schw}=\frac{16 \pi G^{2} M^{2}}{c^{4}}. \label{A_s}
\end{equation}
In the case of a Kerr-Newman black hole, the area is:
\begin{eqnarray}
	A_{\rm KN}&=&4\pi\left(r^{2}_{+}+\frac{a^{2}}{c^{2}}\right) \nonumber\\ &=&4\pi\left[\left(\frac{GM}{c^{2}}+\frac{1}{c^{2}}\sqrt{G^{2}M^{2}-GQ^{2}-a^{2}}\right)^{2}+\frac{a^{2}}{c^{2}}\right].
\label{A_kn}
\end{eqnarray}
Notice that expression (\ref{A_kn}) reduces to (\ref{A_s}) for $a=Q=0$.

The formation of a black holes implies a huge increase of entropy.  Just to compare, a star has an entropy $\sim 20$ orders of magnitude lower than the corresponding black hole of the same mass. This tremendous increase of entropy is related to the loss of all the structure of the original system (e.g. a star) once the black hole is formed. 

The analogy between area and entropy allows to state a set of laws for black holes thermodynamics (Bardeen et al. 1973):
\begin{itemize}
	\item First law (energy conservation): $dM=T_{\rm BH}dS+ \Omega_{+} dJ +\Phi dQ + \delta M$. Here, $\Omega_{+}$ is the angular velocity, $J$ the angular momentum, $Q$ the electric charge, $\Phi$ the electrostatic potential, and $\delta M$ is the contribution to the change in the black hole mass due to the change in the external stationary matter distribution. 
	
	\item Second law (entropy never decreases): In all physical processes involving black holes the total surface area of all the participating black holes can never decrease. 
	
	\item Third law (Nernst's law): The temperature (surface gravity) of a black black hole cannot be zero. Since $T_{\rm BH}=0$ with $A\neq0$ for extremal charged and extremal Kerr black holes, these are thought to be limit cases that cannot be reached in Nature. 
	
	\item Zeroth law (thermal equilibrium): The surface gravity (temperature) is constant over the event horizon of a stationary axially symmetric black hole.    	

\end{itemize}

If a temperature can be associated with black holes, then they should radiate as any other body. The luminosity of a Schwarzschild black hole is:
\begin{equation}
	L_{\rm BH}=4\pi r_{\rm Schw}^{2} \sigma T_{\rm BH}^{4}\sim \frac{16 \pi \sigma \hbar^{4} c^{6}}{(8\pi)^{4} G^{2} M^{2} k^{4}}.
\end{equation}
Here, $\sigma$ is the Stephan-Boltzmann constant. This expression can be written as:
\begin{equation}
	L_{\rm BH}=10^{-17}\left(\frac{M_{\odot}}{M}\right)^{2}\;\;\;{\rm erg\;s}^{-1}.
\end{equation}
 The lifetime of a black hole is:
\begin{equation}
\tau\cong\frac{M}{dM/dt}\sim 2.5 \times 10^{63} \left(\frac{M}{M_{\odot}}\right)^{3}\;\;\;{\rm years}. \label{age}	
\end{equation}
Notice that the black hole heats up as it radiates! This occurs because when the hole radiates, its mass decreases and then according to Eq. (\ref{T}) the temperature must rise.  The black hole then will lose energy and its area will decrease slowly, violating the Second Law of Thermodynamics. However, there is no violation if we consider a {\sl generalised second law}, that always holds: {\sl In any process, the total generalised entropy $S+S_{\rm BH}$ never decreases} (Bekenstein 1973).                      

Unfortunately, many physicists think that entropy and information are the same thing. This confusion seems to come from J. von Neumann, who advised, not without some sarcasm,  Claude Shannon to adopt the expression `entropy' to name the information characterised in the mathematical theory of communications developed by Shannon and Weaver (1949):

\begin{quotation}
You should call it entropy, for two reasons. In the first place your uncertainty function has been used in statistical mechanics under that name, so it already has a name. In the second place, and more important, nobody knows what entropy really is, so in a debate you will always have the advantage.

\begin{flushright}
Floridi (2010), p. 46.
\end{flushright}

\end{quotation}  
     
Shannon's information `entropy', although formally defined by the same expression, is a much more general concept than statistical thermodynamic entropy. Information `entropy' is present whenever there are unknown quantities that can be described only by a probability distribution. When some physicists write about a `Principle of Information Conservation' (e.g. Susskind \& Lindesay 2010), what they really mean is that the entropy of an isolated system in equilibrium should not increase, since it already is at its maximum value. When a black hole accretes matter, however, the entropy increases (they say that ``information is destroyed''). Even if the black hole finally radiates away the whole mass absorbed, the radiation will be thermal, so the entropy of matter will continue to increase. 

As pointed out by Penrose, these considerations do not take into account the entropy of the gravitational field. The state of maximum entropy of this field is gravitational collapse (Penrose 2010). As the black hole evaporates, the entropy of gravitation decreases. Eventually, after the black hole complete evaporation, radiation will be in thermal equilibrium and gravity in a maximally ordered state. After a huge amount of time, the universe might return to a state of minimum overall entropy. Black holes, in this sense, might act as some `entropy regeneration engines', restoring the initial conditions of the universe.    

There is yet another sense of the so-called black hole information paradox, related to the breakdown of predictability of quantum mechanics in presence of black holes. The paradox here appears because of a confusion between ontological and epistemic determinism (see Sect. \ref{Sect3} above). A fundamental postulate of quantum mechanics is that complete description of a system is given by  its wave function up to when the system interacts. The evolution of the wave function is determined by a unitary operator, and unitarity implies epistemic determinism: initial and boundary conditions allow to solve the dynamic equation of the system and the solution is unique. If a system is entangled and one component cross the event horizon, measurements of the second component and knowledge of the initial state will, however, not allow to know the state of the component fallen into the black hole. Epistemic determinism fails for quantum mechanics in presence of black holes. I confess not to see a problem here, since quantum interactions are by themselves already non-unitary. Ontic determinism, the kind that counts, is not in peril here\footnote{See Romero (2012, 2013a) on ontic determinism. }, and epistemic determinism was never part of a full theory of quantum mechanics.  

\section{Inside Black Holes}\label{Sect7}

We have seen that black hole space-times are singular, at least in standard General Relativity.  Moreover, singularity theorems formulated by Penrose (1965) and  Hawking \& Penrose (1970) show that this is an essential feature of black holes. Nevertheless, essential or true singularities should not be interpreted as representations of physical objects of infinite density, infinite pressure, etc. Since the singularities do not belong to the manifold that represents space-time in General Relativity, they simply cannot be described or represented in the framework of such a theory. General Relativity is incomplete in the sense that it cannot provide a full description of the gravitational behaviour of any physical system. True singularities are not within the range of values of the bound variables of the theory: they do not belong to the ontology of a world that can be described with 4-dimensional differential manifolds. Let us see this in more detail (for further discussions see Earman 1995).

A space-time model is said to be singular if the manifold $E$ is {\sl incomplete}. A manifold is incomplete if it contains at least one {\sl inextendible} curve. A curve $\gamma:[0,a)\longrightarrow E$ is inextendible if there is no point $p$ in $E$ such that $\gamma(s)\longrightarrow p$ as $a\longrightarrow s$, i.e. $\gamma$ has no endpoint in $E$. A given space-time model $\left\langle E, \;g_{ab}\right\rangle$ has an {\sl extension} if there is an isometric embedding $\theta: M\longrightarrow E^{\prime}$, where $\left\langle E^{\prime}, g_{ab}^{\prime}\right\rangle$ is another space-time model and $\theta$ is an application onto a proper subset of $E^{\prime}$. A {\em singular} space-time model contains a curve $\gamma$ that is inextendible in the sense given above. Singular space-times are said to contain singularities, but this is an abuse of language: singularities are not `things' in space-time, but a pathological feature of some solutions of the fundamental equations of the theory.  

Singularity theorems can be proved from pure geometrical properties of the space-time model (Clarke 1993). The most important of these theorems is due to Hawking and Penrose (1970):\\

{\bf Theorem.} Let $\left\langle E,\;g_{ab}\right\rangle$ be a time-oriented space-time satisfying the following conditions:
\begin{enumerate}
	\item $R_{ab}V^{a}V^{b}\geq 0$ for any non space-like $V^{a}$\footnote{$R_{ab}$ is the Ricci tensor obtained by contraction of the curvature tensor of the manifold $E$.}.
	\item Time-like and null generic conditions are fulfilled.
	\item There are no closed time-like curves.
	\item At least one of the following conditions holds
	
\begin{itemize}
	\item a.  There exists a compact\footnote{A space is said to be compact if whenever one takes an infinite number of "steps" in the space, eventually one must get arbitrarily close to some other point of the space. Thus, whereas disks and spheres are compact, infinite lines and planes are not, nor is a disk or a sphere with a missing point. In the case of an infinite line or plane, one can set off making equal steps in any direction without approaching any point, so that neither space is compact. In the case of a disk or sphere with a missing point, one can move toward the missing point without approaching any point within the space. More formally,  a topological space is compact if, whenever a collection of open sets covers the space, some sub-collection consisting only of finitely many open sets also covers the space. A topological space is called compact if each of its open covers has a finite sub-cover. Otherwise it is called non-compact. Compactness, when defined in this manner, often allows one to take information that is known locally -- in a neighbourhood of each point of the space -- and to extend it to information that holds globally throughout the space.} achronal set\footnote{A set of points in a space-time with no two points of the set having time-like separation. } without edge.
	\item b. There exists a trapped surface.
	\item c. There is a $p\in E$ such that the expansion of the future (or past) directed null geodesics through $p$ becomes negative along each of the geodesics.  
\end{itemize}
\end{enumerate}
 
Then, $\left\langle E,\;g_{ab}\right\rangle$ contains at least one incomplete time-like or null geodesic. \\

If the theorem has to be applied to the physical world, the hypothesis must be supported by empirical evidence. Condition 1 will be satisfied if the energy-momentum $T^{ab}$ satisfies the so-called {\em strong energy condition}: $T_{ab}V^{a}V^{b}\geq -(1/2)T^{a}_{a}$, for any time-like vector $V^{a}$. If the energy-momentum is diagonal, the strong energy condition can be written as $\rho+3 p\geq 0$ and $\rho + p\geq 0$, with $\rho$ the energy density and $p$ the pressure. Condition 2 requires that any time-like or null geodesic experiences a tidal force at some point in its history. Condition 4a requires that, at least at one time, the universe is closed and the compact slice that corresponds to such a time is not intersected more than once by a future directed time-like curve. The trapped surfaces mentioned in 4b refer to surfaces inside the horizons, from where congruences focus all light rays on the singularity.  Condition 4c requires that the universe is collapsing in the past or the future. 
 
I insist, the theorem is purely geometric, no physical law is invoked. Theorems of this type are a consequence of the gravitational focusing of congruences. 

Singularity theorems are not theorems that imply physical existence, under some conditions, of space-time singularities. Material existence cannot be formally implied. Existence theorems imply that under certain assumptions there are functions that satisfy a given equation, or that some concepts can be formed in accordance with some explicit syntactic rules. Theorems of this kind state the possibilities and limits of some formal system or language. The conclusion of the theorems, although not obvious in many occasions, are always a necessary consequence of the assumptions made. 

In the case of singularity theorems of classical field theories like General Relativity, what is implied is that under some assumptions the solutions of the equations of the theory are defective beyond repair. The correct interpretation of these theorems is that they point out the {\em incompleteness} of the theory: there are statements that cannot be made within the theory. In this sense (and only in this sense), the theorems are like G\"odel's famous theorems of mathematical logic\footnote{G\"odel's incompleteness theorems are two theorems of mathematical logic that establish inherent limitations of all but the most trivial axiomatic systems capable of doing arithmetic. The first theorem states that any effectively generated theory capable of expressing elementary arithmetic cannot be both consistent and complete (G\"odel 1931). The second incompleteness 
theorem, shows that within such
a system, it cannot be 
demonstrated its own 
consistency.}.  

To interpret the singularity theorems as theorems about the existence of certain space-time models is wrong. Using elementary second order logic is trivial to show that there cannot be non-predicable objects (singularities) in the theory (Romero 2013b). If there were a non-predicable object in the theory,
\begin{equation}
	\left(\exists x\right)_{E} \; \left(\forall P\right) \sim Px, \label{P}
\end{equation}
where the quantification over properties in unrestricted. The existential quantification $\left(\exists x\right)_{E}$, on the other hand, means

$$\left(\exists x\right)_{E} \equiv \left(\exists x\right) \wedge \left( x\in E\right).$$ 

Let us call $P_{1}$ the property `$x\in E$'. Then, formula (\ref{P}) reads:
\begin{equation}
\left(\exists x\right)	\left(\forall P\right) (\sim Px \; \wedge P_{1}x ), \label{P1}
\end{equation}
which is a contradiction, i.e. it is false for any value of $x$. 

I conclude that there are no singularities nor singular space-times. There is just a theory with a restricted range of applicability.

The reification of singularities can lead to accept an incredible ontology. We read, for instance, in a book on foundations of General Relativity: 

\begin{quotation}
\noindent [...] a physically realistic space-time {\em must} contain such singularities. [...] there exist causal, inextendible geodesics which are incomplete. [...] If a geodesic cannot be extended to a complete one (i.e. if its future endless continuation or its past endless continuation is of finite length), then either the particle suddenly ceases to exist or the particle suddenly springs into existence. In either case this can only happen if space-time admits a ``singularity'' at the end (or the beginning) of the history of the particle.

\begin{flushright}
Kriele (1999), p. 383.
\end{flushright}
 
\end{quotation}
\vspace{0.5cm}

This statement and many similar ones found in the literature commit the elementary fallacy of confusing a model with the object being modelled. Space-time does not contain singularities. Some of our space-time models are singular. It is this incomplete character of the theory that prompt us to go beyond General Relativity in order to get a more comprehensive view of the gravitational phenomena. As it was very clear to Einstein, his general theory breaks down when the gravitational field of quantum objects starts to affect space-time. 

Another interesting feature of black hole interiors is the existence, according to the unperturbed theory, of a region with closed time-like curves (CTCs) in Kerr and Kerr-Newman black holes. This is the region interior to the second horizon; chronology violation is generated by the tilt of the light cones around the rotation axis in this part of space-time (e.g. Andréka, Ni\'emeti, \& W\"uthrich 2008).  The interior event horizon is also a Cauchy
horizon -- a null hypersurface which is the boundary
of the future domain of dependence for Cauchy data
of the collapse problem. It results impossible to predict the evolution of any system inside the Cauchy horizons; they are an indication of the breaking of predictability in the theory. These horizons, however,  exhibit highly
pathological behaviour; small time-dependent perturbations
originating outside the black hole undergo an infinite
gravitational blueshift as they evolve towards the horizon. This blueshift of infalling radiation
gave the first indications that these solutions may
not describe the generic internal structure of real black holes. Simpson \& Penrose (1973)
pointed this out more than 40 years ago, and
since then linear perturbations have been analysed in detail. 
Poisson \& Israel (1990)
showed that a scalar curvature singularity forms
along the Cauchy horizon of a charged, spherical black hole in a
simplified model. This singularity is characterised by
the exponential divergence of the mass function with advanced
time. The key ingredient producing this 
growth of curvature is the blueshifted radiation flux
along the inner horizon (see also Gnedin \& Gnedin 1993 and Brady 1999 for a review). Since then, the result was generalised to Kerr black holes (e.g. Brady \& Chambers 1996, Hamilton \& Polhemus 2011). These, and other results about the instability of the Kerr black hole interior, suggest that CTCs actually do not occur inside astrophysical black holes.

\section{Black Holes and the Future of the Universe}\label{Sect8}

According to Eq. (\ref{age}), an isolated black hole with $M=10$ $M_{\odot}$ would have a lifetime of more than $10^{66}$ yr. This is 56 orders of magnitude longer than the age of the universe. However, if the mass of the black hole is small, then it could evaporate within the Hubble time. A primordial black hole, created by extremely energetic collisions short after the Big Bang, should have a mass of at least $10^{15}$ g in order to exist today. Less massive black holes must have already evaporated. What happens when a black hole losses its mass so it cannot sustain an event horizon anymore? As the black hole evaporates, its temperature raises. When it is cold, it radiates low energy photons. When the temperature increases, more and more energetic particles will be emitted. At some point gamma rays would be produced. If there is a population of primordial black holes, their radiation should contribute to the diffuse gamma-ray background. This background seems to be dominated by the contribution of unresolved Active Galactic Nuclei and current observations indicate that if there were primordial black holes their mass density should be less than $10^{-8}\;\Omega$, where $\Omega$ is the cosmological density parameter ($\sim 1$). After producing gamma rays, the mini black hole would produce leptons, quarks, and super-symmetric particles, if they exist. At the end, the black hole would have a quantum size and the final remnant will depend on the details of how gravity behaves at Planck scales. The final product might be a stable, microscopic object with a mass close to the Planck mass. Such particles might contribute to the dark matter present in the Galaxy and in other galaxies and clusters. The cross-section of black hole relics is extremely small: $10^{-66}$ cm$^{2}$ (Frolov and Novikov 1998), hence they would be basically non-interacting particles. A different possibility, advocated by Hawking (1974), is that, as a result of the evaporation nothing is left behind: all the energy is radiated. 

Independently of the problem of mini black hole relics, it is clear that the fate of stellar-mass and supermassive black holes is related to fate of the whole universe. In an ever expanding universe or in an accelerating universe as it seems to be our actual universe, the fate of the black holes will depend on the acceleration rate. The local physics of the black hole is related to the cosmic expansion through the cosmological scale factor $a(t)$ (Faraoni \& Jacques 2007). A Schwarzschild black hole embedded in a Friedmann-Lemaitre-Robertson-Walker (FLRW) universe can be represented by a generalisation of the McVittie metric (e.g. Gao et al. 2008):    

\begin{equation}
	ds^{2}=\frac{\left[1-\frac{2G M(t)}{a(t)c^{2}r}\right]^{2}}{\left[1+\frac{2G M(t)}{a(t)c^{2}r}\right]^{2}} c^{2}dt^{2}-a(t)^{2}\left[1+\frac{2G M(t)}{a(t)c^{2}r}\right]^{4} (dr^{2} +r^{2}d\Omega^{2}). \label{cosmicBH}
\end{equation}

Assuming that $M(t)=M_{0} a(t)$, with $M_0$ a constant, the above metric can be used to study the evolution of the black hole as the universe expands. If the equation of state for the cosmic fluid is given by $P=\omega\rho c^{2}$, with $\omega$ constant, then for $\omega<-1$ the universe accelerates its expansion in such a way that the scale factor diverges in a finite time. This time is known as the Big Rip. If $\omega=-1.5$, then the Big Rip will occur in $\sim 35$ Gyr. The event horizon of the black hole and the cosmic apparent horizon will coincide for some time $t<t_{\rm Rip}$ and then the inner region of the black hole would be accesible to all observers. In case of $\omega>-1$ the expansion will continue during an infinite time. Black holes will become more and more isolated. As long as their temperature be higher than that of the Cosmic Microwave Background radiation (CMB), they will accrete photons and increase their mass. When, because of the expansion, the CMB temperature falls below that of the black holes, they will start to evaporate. On the very long run, all black holes will disappear. If massive particles decay into photons on such long timescales, the final state of the universe will be that of a dilute photon gas. Cosmic time will cease to make any sense for such a state of the universe, since whatever exist will be on a null surface. Without time, there will be nothing else to happen. Penrose (2010), however, has suggested that  a countable sequence of open FLRW space-times, each representing a big bang followed by an infinite future expansion might occur, since the past conformal boundary of one copy of FLRW space-time can be ``attached'' to the future conformal boundary of another, after an appropriate conformal rescaling.  Since bosons obey the laws of conformally invariant quantum theory, they will behave in the same way in the rescaled sections of the cyclical universe.  For bosons, the boundary between different cycles is not a boundary at all, but just a space-like surface that can be passed across like any other. Fermions, on the other hand, remain confined to each cycle, where they are generated and decay. Most of the fermions might be converted into radiation in black holes. If this is correct, black holes would then be the key to the regeneration of the universe.

\section{Closing Remarks}\label{Sect9}

In this chapter I have overviewed some philosophical problems related to black holes. The interface between black hole physics and philosophy remains mostly unexplored, and the list of topics I have selected is by no means exhaustive. The study of black holes can be a very powerful tool to shed light on many other philosophical issues in the philosophy of science and even in General Relativity. Evolving black holes, black hole dependence of the asymptotic behaviour of space-time, the nature of inertia, the energy of the gravitational field, quantum effects in the near horizon region, turbulent space-time during black hole mergers, the classical characterisation of the gravitational field, and regular black hole interiors are all physical topics that have philosophical significance. In black holes our current representations of space, time, and gravity are pushed to their very limits. The exploration of such limits can pave the way to new discoveries about the world and our ways of representing it. Discoveries in both science and philosophy.       

\section*{Acknowledgments}
I thank Mario Bunge, Daniela P\'erez, Gabriela Vila,  Federico Lopez Armengol, and Santiago Perez Bergliaffa for illuminating discussions on science and black holes. I am also very grateful to Florencia Vieyro for help with the figures.  My work has been partially supported by  the Argentinian Agency ANPCyT (PICT 2012-00878) and the Spanish MINECO under grant AYA2013-47447-C3-1-P.


\begin{thebibliography}{}

\bibitem{Andreka} Andréka, H., Németi, I., \& W\"uthrich, C. 2008, {\sl Class. Quantum Grav.}, 40, 1809-1823.

\bibitem{Bardeen} 
Bardeen, J. M., Carter, B., \&  Hawking, S. W. 1973, {\sl Communications in Mathematical Physics}, 31 (2), 161–170.

\bibitem{Bekenstein1973} Bekenstein, J.D. 1973, {\sl Phys. Rev. D}, 7, 2333-2346.

\bibitem{Boltzmann}
 Boltzmann, L. 1895,  {\em Nature}, 51, 413-415.

\bibitem{Brady-Chambers} Brady, P.R., \& Chambers, C.M. 1995, {\sl Phys. Rev. D}, 51, 4177-4186.

\bibitem{Brady} Brady, P.R. 1999, {\sl Progress of Theoretical Physics Supplement}, 136, 29-44.

\bibitem{Bunge1977}
Bunge, M. 1977, Ontology I: The Furniture of the World, Kluwer, Dordrecht.

\bibitem{Burbury1}
Burbury, S.H. 1894,  {\em Nature},  51, 78-79.

\bibitem{Burbury2}
Burbury, S.H. 1895,  {\em Nature},  51, 320-320.

\bibitem{Came-07-Springer}
Camenzind, M. 2007, Compact objects in Astrophysics : White Dwarfs, Neutron Stars and Black Holes, Springer, Berlin.

\bibitem{Crisp-03-Oxford} Crisp, T. 2003, Presentism. In: M. J. Loux \& D. W. Zimmerman (Eds.), {\em The Oxford Handbook of Methaphysics}, pp. 211-245, Oxford University Press, Oxford.

\bibitem{Earman}  Earman, J., 1986,  A Primer on Determinism, Reidel, Dordrecht.

\bibitem{Earman1993} Earman, J., \& Norton, J. 1993, {\sl Philos. Sci.}, 60, 22-42.

\bibitem{Earman1995} Earman, J., 1995, Bangs, Crunches, Whimpers, and Shrieks: Singularities and Acausalities in Relativistic Spacetimes,  Oxford University Press, New York.

\bibitem{Eddington}
Eddington, A.S. 1931,  {\em Nature},  127, 447-453.

\bibitem{Falcke-2011-IAU}
Falcke, H., Markoff S., Bower G. C., Gammie, C. F., Moscibrodzka, M. \& Maitra, D. 2011. The Jet in the Galactic Center: An Ideal Laboratory for Magnetohydrodynamics and General Relativity. In:  G. E. Romero, R. A. Sunyaev $\&$ T. Belloni (Eds.), {\em  Jets at all Scales}, Proceedings of the International Astronomical Union, IAU Symposium, Volume 275, 68-76, Cambridge University Press, Cambridge.

\bibitem{Faraoni}Faraoni, V., \& Jacques, A. 2007, {\sl Phys. Rev. D}, 76, id. 063510.

\bibitem{Floridi} Floridi, L. 2010, Information. A Very Short Introduction, Oxford University Press, Oxford.

\bibitem{F-N} Frolov, V.P., \& and Novikov, I.D. 1998, Black Hole Physics, Kluwer, Dordrecht.

\bibitem{Gao} Gao, C., et al. 2008, {\sl Phys. Rev. D}, 78, id. 024008. 

\bibitem{Gnedin-Gnedin} Gnedin, M. L., \& Gnedin, N. Y. 1993, {\sl Class. Quantum Grav.}, 10, 1083-1102.

\bibitem{Godel}
G\"odel, K.  1931, {\sl Monatshefte f\"ur Mathematik und Physik}, 38, 173-198.

\bibitem{Hamilton} Hamilton, A.J.S., \& Polhemus, G. 2011, {\sl Phys. Rev. D}, 84, id. 124055. 

\bibitem{Hawking-BH2} 
Hawking, S.W. 1971,  {\sl Physical Review Letters},  26 (21), 1344–1346.

\bibitem{Hawking1974} Hawking, S.W. 1974, {\sl Nature}, 248, 30-31.


\bibitem{HawkingEllis1973} Hawking, S.W., and Ellis, G.F.R. 1973, The Large-Scale Structure of Space-Time, Cambridge University Press, Cambridge.

\bibitem{Turing-Machine}
Hopcroft, J., \&  Ullman, J. 1979, Introduction to Automata Theory, Languages, and Computation (1st ed.),  Addison–Wesley, Reading Mass.

\bibitem{Hogarth} Hogarth, M. 1994, Non-Turing computers and non-Turing computability. In: Hull, D., Forbes, M., Burian,
R. (Eds.), {\em Proceedings of the Biennial Meeting of the Philosophy of Science Association 1994}, pp. 126–138, University of Chicago Press, Chicago.

\bibitem{Hoyng-2006-PP}
Hoyng, S. 2006, Relativistic Astrophysics and Cosmology: A Primer, Springer, Berlin.

\bibitem{Kriele}
Kriele, M. 1999, Spacetime: Foundations of General Relativity and Differential Geometry, Springer, Berlin-Heidelberg-New York.

\bibitem{Loschmidt} 
Loschmidt, J. 1876,  {\em Wiener Berichte},  73, 128-142.

\bibitem{Mink} Minkowski, H., 1908, Lecture ``Raum und Zeit, 80th Versammlung Deutscher Naturforscher (K$\ddot{o}$ln, 1908)'', {\em Physikalische Zeitschrift}, 10, 75-88 (1909).

\bibitem{Mira-1998-al} 
Mirabel, I.F., Dhawan, V., Chaty, S., Rodriguez, L. F., Marti, J.,Robinson, C. R.,Swank, J., Geballe, T. 1998, {\sl Astronomy  \& Astrophysics}, 330, L9-L12.

\bibitem{N-D2006} N\'emeti, I., \&  David, G. 2006, {\sl Applied Mathematics and Computation}, 178, 118-142.

\bibitem{N-A2006} N\'emeti, I., \&  Andr\'eka, H. 2006, New Physics and Hypercomputation. In:
J. Wiedermann et al. (Eds.), {\em SOFSEM 2006}, LNCS 3831, p. 63, 2006.

\bibitem{Paredes-2009} 
Paredes, J. M. 2009, Black Holes in the Galaxy. In:  G. E. Romero \& P. Benaglia (Eds.),  {\em Compact Objects and their Emission}, Argentinian Astronomical Society Book Series, Volume 1, 91-121. 

\bibitem{Penrose} 
Penrose, R. 1979, Singularities and Time-Asymmetry. In: S.W. Hawking \& W. Israel (Eds.), {\em General Relativity: An Einstein Centennial}, Cambridge University Press, Cambridge, p.581, 1979. 

\bibitem{Penrose2010} Penrose, R. 2010, Cycles of Time, Vintage Books, London.

\bibitem{Piran-2007-al} 
Piran, T. ,\& Fan, Y. 2007, {\sl Philosophical Transactions of the Royal Society A}, 365, 1151-1162.

\bibitem{Poisson-Israel} Poisson, E., \& Israel, W. 1990, {\sl Phys. Rev. D}, 41, 1796-1809.

\bibitem{Price} Price, H. 2004. In: {\em Contemporary Debates in Philosophy of Science}, C. Hitchcock (ed.), Blackwell, Singapore, p. 21.

\bibitem{Punsly2001} Punsly, B. 2001, Black Hole Gravitohydromagnetics, Springer, Berlin.

\bibitem{Rea-03-Oxford} Rea, M. C. 2003, Four-Dimensionalism. In: M. J. Loux \& D. W. Zimmerman (Eds.), {\em The Oxford Handbook of Methaphysics}, pp. 246-80, Oxford University Press, Oxford.

\bibitem{Romero2012-par} Romero, G.E.  2012, {\sl Foundations of Science}, 17, 291-299.

\bibitem{Romero2013-ele } Romero, G.E. 2013a, {\sl  Foundations of Science}, 18, 139-148.

\bibitem{Romero2013-adv } Romero, G.E. 2013b, {\sl  Foundations of Science}, 18, 297-306.

\bibitem{Romero2014a-superT } Romero, G.E. 2014a, {\sl  Foundations of Science}, 19, 209-216.

\bibitem{Romero2014-ont-GR} Romero, G.E. 2014b, in M. Novello, S.E. Perez Bergliaffa, {\em Gravitation 
and Cosmology}, Cambridge, Cambridge Scientific Publishers, Ltd., in press. 

\bibitem{Romero-Perez-2011} Romero, G.E., \& P\'erez, D. 2011,  {\sl Int. J. Modern Phys. D},   20, 2831-2838.

\bibitem{Romero-Perez-2012}  Romero, G. E., Thomas, R., \& P\'erez, D. 2012, {\sl Int. J. Theor. Phys.}, 51, 925-942.

\bibitem{Romero2014book} Romero, G.E., \& Vila, G.S. 2014, Introduction to Black Hole Astrophysics, Springer, Heidelberg.

\bibitem{Romero2014-BH-pres} Romero, G.E., \& P\'erez, D. 2014, {\sl European Journal for Philosophy of Science}, in press, DOI 10.1007/s13194-014-0085-6.

\bibitem{Shannon} Shannon, C.E., \& Weaver, W. 1949, The Mathematical Theory of Communications, University of Illinois Press, Urbana Il. 

\bibitem{Simpson-Penrose} Simpson, M, \& Penrose, R. 1973, {\sl Int. J. Theor. Phys.}, 7, 183-197.

\bibitem{Susskind} Susskind, L., \& Lindesay, J. 2010, An Introduction to Black Holes, Information, and the String Theory Revolution, World Scientific, Singapore. 

\bibitem{Turing} Turing, A. 1936, {\sl Proceedings of the London Mathematical Society, Series 2}, 42, 230–265.

\bibitem{Wald1984} Wald, R.M. 1984, General Relativity, The University of Chicago Press, Chicago.

\bibitem{Weingard} Weingard, R. 1979,  {\sl Synthese}, 42, 191-219. 

\bibitem{Woosley-1993-PP}
Woosley, S. E. 1993, {\sl ApJ}, 405, 273-277.

\end{thebibliography}
\end{document}